\documentclass[aps,prb,superscriptaddress]{revtex4}
\usepackage{times}
\usepackage{epsfig}
\usepackage{graphicx}
\usepackage{amsmath}

\begin{document}
\title{Spin-orbit coupling induced interference in quantum corrals}
\author{Jamie D. Walls}\email{jwalls@fas.harvard.edu}\affiliation{Department of Chemistry and Chemical Biology,
 Harvard University, Cambridge, MA 02138}
 \author{Eric J. Heller}\email{heller@physics.harvard.edu}\affiliation{Department of Chemistry and Chemical Biology,
 Harvard University, Cambridge, MA 02138}\affiliation{Department of
 Physics, Harvard University, Cambridge, MA 02138}
\date{\today}
\begin{abstract}
Lack of inversion symmetry at a metallic surface can lead to an
observable spin-orbit interaction.  For certain metal surfaces, such
as the Au(111) surface, the experimentally observed spin-orbit
coupling results in spin rotation lengths on the order of tens of
nanometers, which is the typical length scale associated with
quantum corral structures formed on metal surfaces. In this work,
multiple scattering theory is used to calculate the local density of
states ($LDOS$) of quantum corral structures comprised of
nonmagnetic adatoms in the presence of spin-orbit coupling.
 Contrary to previous theoretical
predictions, spin-orbit coupling induced modulations are observed in
the theoretical $LDOS$, which should be observable using scanning
tunneling microscopy.
\end{abstract}
\maketitle
 In the presence of time reversal symmetry
$[E(k,\uparrow)=E(-k,\downarrow)]$ and spatial inversion symmetry
$[E(k,\uparrow)=E(-k,\uparrow)]$, no spin splitting can exist since
$E(k,\uparrow)=E(k,\downarrow)$. At a metal surface, however,
spatial inversion symmetry is violated, and a spin splitting can
therefore occur, i.e., $E(k,\uparrow)\neq E(k,\downarrow)$.   The
spin-orbit coupling in surface states was first observed by LaShell
et al.\cite{LaShell96} on the Au(111) surface using photoemission
spectroscopy.  The form of the spin-orbit interaction was found to
be similar to the Rashba spin-orbit coupling\cite{Bychkov84}, which
has been heavily studied in semiconductor heterostructures and
quantum wells.  Additional experimental\cite{Nicolay01,Henk03} and
theoretical\cite{Petersen00,Reinert03,Bihlmayer06} evidence have
confirmed the presence of significant spin-orbit coupling on the
Au(111) surface.  Although such a spin-splitting should, in
principle, occur on all surfaces, the magnitude of the spin
splitting depends very strongly on the nature of the surface. For
instance, spin-orbit coupling has never been observed on either the
Ag(111) or the Cu(111) surfaces.  This is due to the fact that the
magnitude of the spin-orbit coupling is determined largely by the
atomic spin-orbit coupling and the gradient of the surface state
wave function at the nucleus\cite{Bihlmayer06}; theoretical
calculations, which accurately predict the observed spin-orbit
coupling on the Au(111) surface, predict the spin-orbit coupling on
the Ag(111) to be a factor of 20 smaller than the spin-orbit
coupling on the Au(111) surface\cite{Reinert03,Bihlmayer06}, well
outside the range of current experimental observation. In addition
to the Au(111) surface, photoemission experiments have discovered a
variety of other metallic systems with spin-orbit coupling, such as
on the Bi surfaces\cite{Koroteev04}, which exhibit an even larger
spin-orbit coupling than that found on Au(111).

Although most experimental observations of spin-orbit coupling in
surface states are from photoemission spectroscopy, scanning
tunneling microscopy (STM) has been used to observe spin-orbit
interference in a magnetic sample\cite{Bode03} and in nonmagnetic
systems with very strong spin-orbit coupling, such as on the Bi(110)
surface\cite{Pascual04} and in Bi/Ag(111) and Pb/Ag(111) surface
alloys\cite{Ast07}.  However, previous theoretical
work\cite{Petersen00} has argued that scanning tunneling microscopy
(STM) could not be used to observe the spin-orbit coupling in
surface states;  this argument was based on the assumption that the
trajectories which interfere at the site of the STM tip are all
one-dimensional in nature.  Such trajectories do not undergo any net
spin rotation, which results in the same standing wave pattern found
in the absence of spin-orbit coupling.  While the above argument is
certainly true for the case of scattering from a single nonmagnetic
adatom, trajectories involving multiple scatterers will undergo a
net spin rotation, which will lead to spin-orbit induced modulations
of the local density of states ($LDOS$), which should be observable
using STM.

Multiple scattering trajectories have been
shown\cite{Heller94,Fiete01,Fiete03} to be important in
understanding the standing wave patterns observed in the $LDOS$ for
step edges\cite{Crommie93}, for quantum corrals\cite{Crommie93a}
formed by placing adatoms atop a noble metal surface, and for
quantum mirages generated by a magnetic adatom placed inside a
quantum corral\cite{Manoharan00,Fiete01}.  Previous experimental
work has been conducted for quantum corrals on either the Cu(111)
surface\cite{Crommie93a,Manoharan00} or on the Ag(111)
surface\cite{Kliewer00}, where the neglect of spin-orbit coupling,
as stated above, is completely justified\cite{Nicolay01,Reinert03}.
  However, this would not be the case for quantum corrals formed on
the Au(111) surface.  In this work, multiple scattering theory in
the presence of spin-orbit coupling is used to calculate the
expected change in the $LDOS$ for quantum corrals formed on surfaces
with significant spin-orbit coupling, such as Au(111). Numerical
calculations performed for both a circular and a stadium quantum
corral formed from nonmagnetic adatoms demonstrate that spin-orbit
coupling can lead to observable changes in the $LDOS$.
Understanding the effects of spin-orbit induced interference on
metal surfaces will be important if such systems are to be used for
future spintronics applications.

\begin{figure}
\includegraphics*[height=10.7cm,width = 15.7cm]{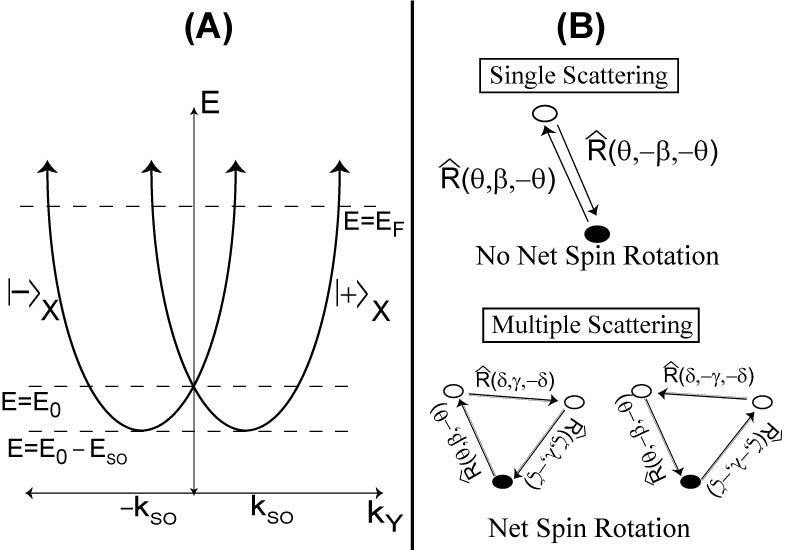}
\caption{(A)  The dispersion curve projected along $k_{X}=0$ in the
presence of spin-orbit coupling.  The normal parabolic dispersion
relation has been split into two parabolic curves, centered about
$k_{Y}=\pm k_{\text{SO}}$, with the band edge occurring at an energy
of $E_{0}-E_{\text{SO}}$. Note that the spin state is
$|+(0)\rangle\equiv |+\rangle_{X}$ for the parabolic band centered
at $k_{Y}=+k_{\text{SO}}$ and $|-(0)\rangle\equiv |-\rangle_{X}$ for
the parabolic band centered at $k_{Y}=-k_{\text{SO}}$.  (B)  For a
single scattering trajectory, spin-orbit coupling cannot generate a
net spin rotation since
$\widehat{R}(\theta,\beta,-\theta)\widehat{R}(\theta,-\beta,-\theta)=\widehat{1}$.
However, for non-collinear multiple scattering trajectories, a net
spin rotation can occur since
$\widehat{R}(\theta,\pm\beta,-\theta)\widehat{R}(\delta,\pm\gamma,-\delta)\widehat{R}(\zeta,\pm\lambda,-\zeta)\neq
\widehat{1}$.
 }\label{fig:fig1}
\end{figure}

 The
effective Hamiltonian for a surface state in the presence of the
Rashba spin-orbit interaction is given by:
\begin{eqnarray}
\widehat{H}=\frac{\widehat{P}_{X}^{2}}{2m^{*}}+\frac{\widehat{P}_{Y}^{2}}{2m^{*}}-\frac{\alpha}{\hbar}\left(\widehat{P}_{Y}\widehat{\sigma}_{X}-\widehat{P}_{X}\widehat{\sigma}_{Y}\right)+E_{0}
\end{eqnarray}
where $m^{*}$ is the effective mass, $\alpha$ is the spin-orbit
coupling strength, and $E_{0}$ is an energy offset arising from the
confinement of the electron to the surface. The eigenstates of
$\widehat{H}$ with energy $E$ are given by
$\Psi_{1}(\vec{r})=\exp(\text{i}\vec{k}_{1}(E)\cdot\vec{r})|+(\phi)\rangle$
and
$\Psi_{2}(\vec{r})=\exp(\text{i}\vec{k}_{2}(E)\cdot\vec{r})|-(\phi)\rangle$,
where the spin quantization axis for $|\pm(\phi)\rangle$ depends
upon the momentum vector,
$\vec{k}_{1(2)}(E)=k_{1(2)}(E)\left(\cos(\phi)\widehat{x}+\sin(\phi)\widehat{y}\right)$,
where $k_{1}(E)=k_{\text{SO}}+\overline{k}(E)$ and
$k_{2}(E)=-k_{\text{SO}}+\overline{k}(E)$, with
$k_{\text{SO}}=\frac{m^{*}\alpha}{\hbar^2}$ and
$\overline{k}(E)=\sqrt{\left(k_{\text{SO}}\right)^{2}+\frac{2m^{*}(E-E_{0})}{\hbar^{2}}}$
(for convenience, the energy dependence of $k_{1}$, $k_{2}$, and
$\overline{k}$ will not be explicitly written from now on). For a
given value of $\phi$, the spin states are
$|\pm(\phi)\rangle=(\sqrt{2})^{-1}\left(|+\rangle_{Z}\pm\exp(-\text{i}\phi)|-\rangle_{Z}\right)$,
where $|\pm\rangle_{Z}$ are eigenstates of $\widehat{\sigma}_{Z}$.
Due to spin-orbit coupling, the dispersion relation,
$E(\vec{k})=\frac{\hbar^{2}|\vec{k}|^{2}}{2m^{*}}\mp
\alpha|\vec{k}|+E_{0}$, consists of two parabolic bands centered
about $\pm k_{\text{SO}}$ with the bottom of the bands occurring at
an energy $E_{0}-E_{\text{SO}}$ (where
$E_{\text{SO}}=\frac{\hbar^{2}k^{2}_{\text{SO}}}{2m^*}$) instead of
at energy $E_{0}$.  The dispersion relation is plotted in Figure
\ref{fig:fig1}(A) for $k_{X}=0$, where the spin states are
$|+(0)\rangle\equiv |+\rangle_{X}$ for the band centered at
$k_{Y}=k_{\text{SO}}$ and $|-(0)\rangle\equiv |-\rangle_{X}$ for the
band centered at $k_{Y}=-k_{\text{SO}}$.  The full two-dimensional
dispersion curve in the $k_{X}-k_{Y}$ plane can be found by simply
rotating the dispersion curve in Fig. \ref{fig:fig1}(A) using
$\exp\left(-\text{i}\theta/\hbar\widehat{L}_{Z}\right)\exp\left(-\text{i}\theta/2\widehat{\sigma}_{Z}\right)$,
where $\theta\in[0,\pi)$ and $\widehat{L}_{Z}$ is the
$\widehat{z}$-component of the angular momentum operator.

In an STM experiment\cite{Tersoff85}, the bias voltage between the
tip and the surface, $V$, can be changed in order to probe the local
density of states at an energy $E_{F}+eV$ (where $E_{F}$ is the
Fermi energy of the metal) by measuring the local conductance,
$\frac{\text{d}I}{\text{d}V}(E_{F}+eV,\vec{r})$, since
$\frac{\text{d}I}{\text{d}V}(E_{F}+eV,\vec{r})\propto
LDOS(E_{F}+eV,\vec{r})$ where
$LDOS(E,\vec{r})=\sum_{q=\pm}\sum_{\nu}\langle
\vec{r},q|\Psi_{\nu}\rangle\langle
\Psi_{\nu}|\vec{r},q\rangle\delta(E-E_{\nu})$.  Thus in order to
calculate the STM image, the $LDOS(E_{F}+eV,\vec{r})$ must be
determined.  One method of determining the $LDOS(E_{F}+eV,\vec{r})$
is by calculating the Green's function,
$\widehat{G}_{\pm}(\vec{r}_{1},\vec{r}_{2},E_{F}+eV)$, and using the
following relationship:
\begin{eqnarray}
LDOS(E_{F}+eV,\vec{r})&=&\frac{\text{i}}{2\pi}\text{Trace}_{\text{spin}}\left[\widehat{G}_{+}(\vec{r},\vec{r},E_{F}+eV)-\widehat{G}_{-}(\vec{r},\vec{r},E_{F}+eV)\right]
\label{eq:ldosg}
\end{eqnarray}
Thus knowledge of the Green's function can be used to calculate the
expected STM signal.

The free-particle Green's function in the presence of Rashba
spin-orbit coupling and for $E\geq E_{0}-E_{\text{SO}}$ is given
by\cite{Walls06,Csordas06}:
\begin{eqnarray}
\widehat{G}^{0}_{\pm}(\vec{r}_{1},\vec{r}_{2},E)&=&\mp\text{i}\frac{m^*}{4\overline{k}\hbar^2}\left(\begin{array}{cc}G^{D}_{\pm}(E,r_{12})&\pm\text{i}\exp(\text{i}\theta_{12})G^{S}_{\pm}(E,r_{12})\\
\pm\text{i}\exp(-\text{i}\theta_{12})G^{S}_{\pm}(E,r_{12})&G^{D}_{\pm}(E,r_{12})\end{array}\right)
\label{eq:GreenSO}
\end{eqnarray}
where $r_{12}=|\vec{r}_{1}-\vec{r}_{2}|$,
$\exp(\pm\text{i}\theta_{12})=[(\vec{r}_{1}-\vec{r}_{2})\cdot\widehat{y}\pm\text{i}(\vec{r}_{1}-\vec{r}_{2})\cdot
\widehat{x}]/r_{12}$, $G^{D}_{\pm}(E,r_{12})=
k_{1}H_{0}^{\pm}(k_{1}r_{12})+k_{2}H_{0}^{\pm}(k_{2}r_{12})$, and
$G^{S}_{\pm}(E,r_{12})=k_{1}H^{\pm}_{1}(k_{1}r_{12})-k_{2}H^{\pm}_{1}(k_{2}r_{12})$.
Note that $k_{2}\leq 0$ for energies $E_{0}-E_{\text{SO}}\leq
E_{F}+eV\leq E_{0}$, so that in this energy range,
$H_{n}^{\pm}(k_{2}r_{12})=(-1)^{n+1}H_{n}^{\mp}(|k_{2}|r_{12})$ in
Eq. (\ref{eq:GreenSO}).  This results in a change in the
$LDOS(E_{F}+eV,\vec{r})$ when $E_{0}-E_{\text{SO}}\leq E_{F}+eV\leq
E_{0}$: for $E_{F}+eV\geq E_{0}$, the free particle
$LDOS_{\text{free}}(E_{F}+eV,\vec{r})$ is independent of energy and
is given by
$LDOS_{\text{free}}(E_{F}+eV,\vec{r})=m^{*}/(\pi\hbar^2)$, whereas
for $E_{0}-E_{\text{SO}}\leq E_{F}+eV\leq E_{0}$, the
$LDOS(E_{F}+eV,\vec{r})$ is dependent upon $E_{F}+eV$ and is given
by $LDOS_{\text{free}}(E_{F}+eV,\vec{r})=m^{*}/(\pi\hbar^{2})\times
k_{\text{SO}}/\overline{k}$.  This change in the
$LDOS_{\text{free}}$ has been recently reported for STM measurements
on Bi/Ag(111) and Pb/Ag(111) surface alloys\cite{Ast07}.

In order to gain more physical insight into the transport between
$\vec{r}_{1}$ and $\vec{r}_{2}$,
$\widehat{G}^{0}_{\pm}(\vec{r}_{1},\vec{r}_{2},E)$ can be rewritten
in terms of a complex amplitude multiplied by a ``complex''
rotation:
\begin{eqnarray}
\widehat{G}^{0}_{\pm}(\vec{r}_{1},\vec{r}_{2},E)&=&\mp\text{i}\frac{m^{*}}{4\overline{k}\hbar^{2}}bb_{\pm}(E,r_{12})\widehat{R}\left(\frac{\theta_{12}}{2},\pm
z_{\pm}(E,r_{12}),-\frac{\theta_{12}}{2}\right)
\end{eqnarray}
where
$bb_{\pm}(E,r_{12})=\sqrt{[G^{D}_{\pm}(E,r_{12})]^{2}+[G^{S}_{\pm}(E,r_{12})]^{2}}$,
$\widehat{R}(\alpha,\beta,\gamma)=\exp(\text{i}\alpha\widehat{\sigma}_{Z})\exp(\text{i}\beta\widehat{\sigma}_{X})\exp(\text{i}\gamma\widehat{\sigma}_{Z})$
is an arbitrary rotation operator with Euler angles $\alpha,\beta,$
and $\gamma$, and $z_{\pm}(E,r_{12})$ is a complex angle which is
defined by
$\cos[z_{\pm}(E,r_{12})]=G_{\pm}^{D}(E,r_{12})/bb_{\pm}(E,r_{12})$
and
$\sin[z_{\pm}(E,r_{12})]=G_{\pm}^{S}(E,r_{12})/bb_{\pm}(E,r_{12})$.
Note that for a trajectory which goes from $\vec{r}_{2}$ to
$\vec{r}_{1}$ and then back to $\vec{r}_{2}$, no net spin rotation
occurs, since
$\widehat{G}^{0}_{\pm}(\vec{r}_{2},\vec{r}_{1})\widehat{G}^{0}_{\pm}(\vec{r}_{1},\vec{r}_{2})\propto
\left(bb_{\pm}(E,r_{12})\right)^{2}\widehat{R}(\frac{\theta_{12}}{2},\pm
z_{\pm}(E,r_{12}),-\frac{\theta_{12}}{2})\widehat{R}(\frac{\theta_{12}}{2},\mp
z_{\pm}(E,r_{12}),-\frac{\theta_{12}}{2})=\left(bb_{\pm}(E,r_{12})\right)^{2}\widehat{1}$.

In the presence of multiple adatoms, the total Green's function can
be significantly altered from
$\widehat{G}^{0}_{\pm}(\vec{r}_{1},\vec{r}_{2},E)$ due to the
interference between the various multiple scattering trajectories.
The total Green's function in the presence of $N$ nonmagnetic
adatoms/scatterers can be approximated as:
\begin{eqnarray}
\widehat{G}_{\pm}(\vec{r}_{1},\vec{r}_{2},E)&=&\widehat{G}_{\pm}^{0}(\vec{r}_{1},\vec{r}_{2},E)+\sum_{j=1}^{N}\widehat{G}_{\pm}^{0}(\vec{r}_{1},\vec{r}_{j},E)\widehat{s}^{\pm}_{j}\widehat{G}_{\pm}(\vec{r}_{j},\vec{r}_{2},E)
\label{eq:msG}
\end{eqnarray}
where $\widehat{s}^{\pm}_{j}$ is the ``s''-wave scattering
amplitude, which is given by
$\widehat{s}^{\pm}_{j}=\hbar^2/m^{*}[\exp(\pm
2\text{i}\delta_{j}(E))-1]\widehat{1}$, with $\delta_{j}(E)$ being
the scattering phase shift.  In writing Eq. (\ref{eq:msG}), the
scattering length of each adatom was assumed to be much smaller than
$2\pi/\overline{k}$ (justifying the ``s''-wave approximation) and
the spin rotation length, $\pi/k_{\text{SO}}$, which allows one to
associate the same scattering amplitude for both the $k_{1}$ and
$k_{2}$ scattered waves (see for example Eqs. (32)-(33) of Ref.
\cite{Walls06}).  The unknown values of the Green's function at each
scatterer $n$, $\widehat{G}_{\pm}(\vec{r}_{n},\vec{r}_{2},E)$, can
be found by setting $\vec{r}_{1}=\vec{r}_{n}$ to give:
\begin{eqnarray}
\widehat{G}_{\pm}(\vec{r}_{n},\vec{r}_{2},E)&=&\widehat{G}_{\pm}^{0}(\vec{r}_{n},\vec{r}_{2},E)+\sum_{j\neq
n}\widehat{G}_{\pm}^{0}(\vec{r}_{n},\vec{r}_{j},E)\widehat{s}^{\pm}_{j}\widehat{G}_{\pm}(\vec{r}_{j},\vec{r}_{2},E)
\end{eqnarray}
This results in a system of $4N$ equations which can be solved via a
simple matrix inversion.  With knowledge of
$\widehat{G}_{\pm}(\vec{r}_{n},\vec{r}_{2},E)$ for each scatterer
$n$, the total Green's function,
$\widehat{G}_{\pm}(\vec{r}_{1},\vec{r}_{2},E)$ in Eq.
(\ref{eq:msG}), is determined, thus determining the $LDOS$ by using
Eq. (\ref{eq:ldosg}).


Consider first the simple case of a single nonmagnetic adatom placed
atop a metal surface at $\vec{r}_{j}$.  The total Green's function
in this case is given by:
\begin{eqnarray}
\vec{G}_{\pm}(\vec{r}_{1},\vec{r}_{2},E)=\widehat{G}_{\pm}^{0}(\vec{r}_{1},\vec{r}_{2},E)+s^{\pm}_{j}\widehat{G}^{0}_{\pm}(\vec{r}_{1},\vec{r}_{j},E)\widehat{G}_{\pm}^{0}(\vec{r}_{j},\vec{r}_{2},E)
\end{eqnarray}
which results in a change in the $LDOS(E,\vec{r})$ of $\Delta
LDOS(E,\vec{r}_{0})=LDOS(E,\vec{r}_{0})-LDOS_{\text{free}}(E,\vec{r}_{0})=\frac{2}{\pi}\left(\frac{m^{*}}{4\overline{k}\hbar^{2}}\right)^{2}\text{Im}\left[s^{+}_{j}(b_{+}(E,r_{0j}))^{2}\right]$,
which, for $\overline{k}r_{0j}\gg 1$ can be approximated as:
\begin{eqnarray}
\Delta LDOS(E_{F}+eV,\vec{r}_{0})&=&-
\frac{\sqrt{k_{1}k_{2}}}{\overline{k}^{2}r_{0j}}\left(\frac{m^{*}}{\pi\hbar^{2}}\right)^{2}\left(\cos(2\overline{k}r_{0j})\text{Re}[s^{+}_{j}]-\sin(2\overline{k}r_{0j})\text{Im}[s^{+}_{j}]\right)
\end{eqnarray}
for $E_{F}+eV>E_{0}$, and as:
\begin{eqnarray}
\Delta LDOS(E_{F}+eV,\vec{r}_{0})&=&
\frac{\sqrt{k_{1}|k_{2}|}}{\overline{k}^{2}r_{0j}}\left(\frac{m^{*}}{\pi\hbar^{2}}\right)^{2}\left(\cos(2\overline{k}r_{0j})\text{Im}[s^{+}_{j}]+\sin(2\overline{k}r_{0j})\text{Re}[s^{+}_{j}]\right)
\end{eqnarray}
for $E_{0}-E_{\text{SO}}\leq E_{F}+eV\leq E_{0}$.  Therefore, there
exists a change in the $\Delta LDOS$ when $E_{0}-E_{\text{SO}}\leq
E_{F}+eV\leq E_{0}$ due to spin-orbit coupling, which is similar to
the change observed in the $LDOS_{\text{free}}$ described
earlier\cite{Ast07}. For the case of a single nonmagnetic adatom,
this change in the $\Delta LDOS$ would be the only way to detect the
presence of spin-orbit coupling, since the period of the spatial
modulation in the $\Delta LDOS$, $2\overline{k}$, can only be used
to determine the effective energy of the surface state electron,
$\overline{k}\equiv \sqrt{\frac{2m^{*}E_{\text{eff}}}{\hbar^{2}}}$.
Spin-orbit coupling only shifts the effective bottom of the band
from $E_{0}$ to $E_{0}-E_{\text{SO}}$, so measurement of
$\overline{k}$ cannot, by itself, help to determine the presence or
absence of spin-orbit coupling.  The physical reason why spin-orbit
coupling doesn't affect the $LDOS$ in the presence of a single
adatom is that for single scattering paths returning to the STM tip,
no net spin rotation can occur, as shown in Fig. \ref{fig:fig1}(B).
This was the reasoning used to argue that STM couldn't be used to
observe spin-orbit coupling for a surface state\cite{Petersen00}.
However, in the presence of multiple adatoms, mutliple scattering
trajectories can generate a net spin rotation (Fig.
\ref{fig:fig1}(B)), which allows the spin-orbit coupling to affect
the $LDOS$ in a nontrivial manner.  As mentioned earlier, such
multiple scattering trajectories are important in understanding the
observed $LDOS$ in quantum corrals formed atop noble metal
surfaces\cite{Heller94,Fiete01,Fiete03}.

For the calculation of the $LDOS$ on the Au(111) surface, the
following parameters were used\cite{LaShell96}: $m^{*}=0.26m_{e}$
and $E_{F}-E_{0}=0.41$ eV (Ref. \cite{Kevan87}), a spin-orbit
coupling constant of $\alpha=4\times 10^{-11}$ eV-m (which is $10\%$
smaller than the value given in Ref. \cite{Henk03} and $21\%$ larger
than the value given in Ref. \cite{LaShell96}).   These parameters
give a Fermi wavelength of $\lambda_{F}=2\pi/\overline{k}=37.4\AA$
and a spin rotation length of $\pi/k_{\text{SO}}=230.5\AA$.  It
should be noted that this spin rotation length is about an order of
magnitude smaller than the attainable spin-rotation lengths in
semiconductor heterostructures, which is mainly attributable to the
larger effective mass of the surface state electrons.

In the following calculations, all adatoms were modeled as
``black-dots''\cite{Heller94} where $\delta(E)=\text{i}\infty$ due
to inelastic scattering of electrons into the bulk\cite{Crampin96}
(modifications of the theory for treating the adatom scattering as
purely elastic\cite{Harbury96} can also be performed too).  In the
simulations, each nonmagnetic adatom was placed on a hexagonal
lattice at a position
$\vec{r}=\frac{a}{2}\left(\left(b_{1}+b_{2}\right)\widehat{x}+\sqrt{3}\left(b_{1}-b_{2}\right)\widehat{y}\right)$,
where $a=2.885\AA$ for Au(111), and $b_{1}$ and $b_{2}$ are integers
chosen to minimize $|\vec{r}-\vec{r}_{\text{d}}|$, where
$\vec{r}_{\text{d}}$ is the desired location for each adatom.  It
should be mentioned that a hexagonal lattice is a simplified model
of the actual Au(111) surface, which undergoes a herringbone
reconstruction\cite{Chen98}. Such a reconstruction acts like a
superlattice for the surface state electrons and modifies the
electron density; however, such a reconstruction has minimal effect
on the spin-orbit coupling as has been demonstrated by theoretical
calculations\cite{Petersen00,Reinert03,Bihlmayer06} and is not
considered in the following simulations.

In order to illustrate the effect of spin-orbit coupling on the
resulting $\Delta LDOS(E,\vec{r})$, simulations with and without
spin-orbit coupling were performed at slightly different applied
voltages but with the same effective energy, $E_{\text{eff}}$, in
order that both simulations gave the same period in the spatial
oscillation of the $\Delta LDOS(E_{\text{eff}},\vec{r})$ in the
presence of a single adatom,
$2\overline{k}=2\sqrt{2m^{*}E_{\text{eff}}/\hbar^{2}}$. For example,
if $V$ was the applied voltage used in the simulation in the absence
of spin-orbit coupling, then the applied voltage in the presence of
spin-orbit coupling, $V'$, would be given by
$eV'=eV-E_{\text{SO}}=eV-2.7$ meV, with
$E_{\text{eff}}=E_{F}-E_{0}+eV=E_{F}-E_{0}+eV'+E_{\text{SO}}$. In
order to consider only the contributions of spin-orbit to the
$\Delta LDOS$ arising from multiple scattering trajectories,
effective energies, $E_{\text{eff}}\geq E_{\text{SO}}$, were only
considered in order to avoid the intrinsic change in the $\Delta
LDOS$
 when $0\leq E_{\text{eff}}\leq E_{\text{SO}}$.  Note that for the case of the Au(111) surface,
this intrinsic change in the $\Delta LDOS$ should in any case be
unobservable since $E_{\text{SO}}=2.7$ meV is much smaller than the
lifetime broadening\cite{Ast07} of $18$ meV.

\begin{figure}
\includegraphics*[height=8.7cm,width = 7.7cm]{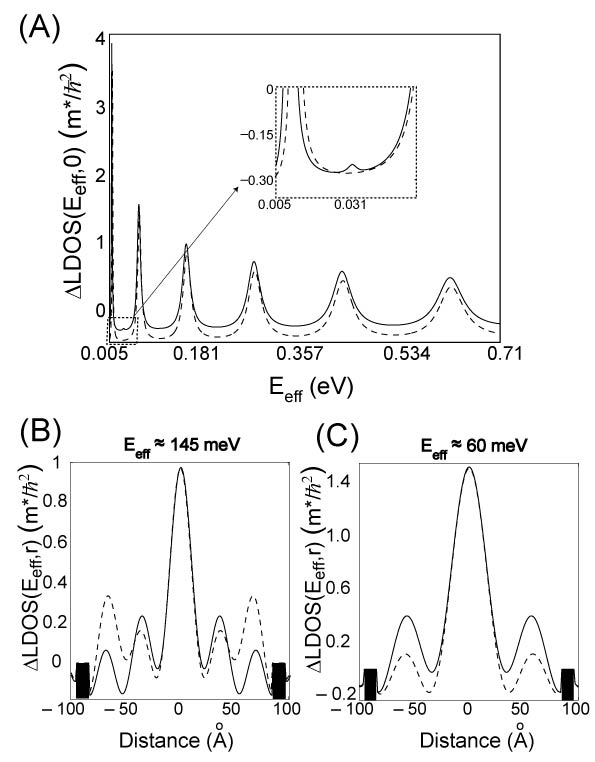}
\caption{The $\Delta LDOS(E,\vec{r})$ for a circular quantum of
radius $88.7\AA$ comprised of 60 nonmagnetic adatoms, which are
modeled as ``black-dot'' scatterers ($\delta (E)=\text{i}\infty$).
In (A), the $\Delta LDOS(E_{\text{eff}},0)$ is plotted at the center
of the corral as a function of the effective energy,
$E_{\text{eff}}$, with (solid curve) and without (dashed curve)
spin-orbit coupling. The dashed curve has been shifted downward for
convenience.  Note that $E_{\text{eff}}$ can be converted into a
bias voltage by using either
$-eV=E_{F}-E_{0}-E_{\text{eff}}\,\text{(dashed curve)}$ or
$-eV'=E_{F}-E_{0}+E_{\text{SO}}-E_{\text{eff}}=E_{F}-E_{0}-E_{\text{eff}}+2.7$
meV (solid curve), where $E_{F}-E_{0}=410$ meV for the Au(111)
surface. The following parameters were used in the simulation:
$m^*=0.26m_{e}$ and $\alpha=4\times 10^{-11}$ eV-m. The peaks in the
$\Delta LDOS(E_{\text{eff}},0)$ occur when $E_{\text{eff}}$ roughly
corresponds to eigenergy for a circular billiard with (solid curve)
and without (dashed curve) spin-orbit coupling. A small peak (shown
in the inset) in the $\Delta LDOS(E_{\text{eff}},0)$ at
$E_{\text{eff}}\approx 31$ meV, corresponds to an eigenstate of the
circular billiard with spin-orbit coupling which is mostly
$J_{1}(k|\vec{r}|)\exp(\pm\text{i}\theta)$ in character, but, due to
spin-orbit coupling, does contain some $J_{0}(k|\vec{r}|)$
character, which can contribute to the $\Delta
LDOS(E_{\text{eff}},0)$. In (B) and (C), profiles of the $\Delta
LDOS(E_{\text{eff}},\vec{r})$ through the quantum corral (the
adatoms are denoted by the black rectangles) with (solid curve) and
without (dashed curve) spin-orbit coupling for (C) the second main
peak in the $\Delta LDOS(E_{\text{eff}},0)$ [$E_{\text{eff}}=60.4$
meV (without spin-orbit coupling) and $E_{\text{eff}}=58.7$ meV
(with spin-orbit coupling)] and for (B) the third main peak in the
$\Delta LDOS(E_{\text{eff}},0)$ [$E_{\text{eff}}=145.8$ meV (without
spin-orbit coupling) and $E_{\text{eff}}=144.1$ meV (with spin-orbit
coupling)].  In both cases, substantial differences in the intensity
of the $\Delta LDOS(E_{\text{eff}},\vec{r})$ are observed, where the
presence of spin-orbit coupling can either enhance the $\Delta
LDOS(E_{\text{eff}},\vec{r})$ (Fig. \ref{fig:fig2}(C)) or decrease
the $\Delta LDOS(E_{\text{eff}},\vec{r})$ (Fig. \ref{fig:fig2}(B)).
 }\label{fig:fig2}
\end{figure}

Simulations were first performed on a circular quantum corral of
radius $88.7\AA$ comprised of sixty nonmagnetic adatoms placed atop
a hexagonal lattice   The calculated $\Delta LDOS(E_{\text{eff}},0)$
at the center of the corral is shown in Figure \ref{fig:fig2}(A) as
a function of $E_{\text{eff}}$ in the presence (solid curve) and in
the absence (dashed curve) of spin-orbit coupling. The $\Delta
LDOS(E_{\text{eff}},0)$ without spin-orbit coupling has been shifted
down for convenience. A very simple ``particle in a box''
model\cite{Crommie93a} can be used to interpret the $\Delta
LDOS(E_{\text{eff}},0)$ in Fig. \ref{fig:fig2}(A): in the absence of
spin-orbit coupling and treating the quantum corral as a circular
billiard with radius $R=88.7\AA$, the peaks in the $\Delta
LDOS(E_{\text{eff}},0)$ mainly occur when $E_{\text{eff}}$ is equal
to an eigenenergy of the circular billiard,
$E_{\text{eff}}=E_{n}=\frac{\hbar^{2}(k_{n,0})^{2}}{2m^{*}}$ where
$k_{n,0}$ is given by the solution to $J_{0}(k_{n,0}R)=0$.  This
simple model predicts the peak locations in the $\Delta
LDOS(E_{\text{eff}},0)$ to within 10 meV for the first four peaks
shown in Fig. \ref{fig:fig2}(A).

A similar model can be applied to the case of a circular billiard
with spin-orbit coupling. In this case the eigenstates can be
written as:\begin{eqnarray}
 \Psi_{n,m}(\vec{r})&\propto&\exp(\text{i}m\theta)\left(\begin{array}{c}J_{m}(k_{1}^{n,m}|\vec{r}|)-\frac{J_{m}(k_{1}^{n,m}R)}{J_{m}(k_{2}^{n,m}R)}J_{m}(k_{2}^{n,m}|\vec{r}|)\nonumber\\
-\text{i}\exp(-\text{i}\theta)\left(J_{m-1}(k^{n,m}_{1}|\vec{r}|)+\frac{J_{m}(k_{1}^{n,m}R)}{J_{m}(k_{2}^{n,m}R)}J_{m-1}(k_{2}^{n,m}|\vec{r}|)\right)\end{array}\right)
 \label{eq:spincirc}
 \end{eqnarray}
 which have an effective energy (shifted by $-E_{\text{SO}}$ for comparison to the simulations without spin-orbit coupling) given by
 $E_{\text{eff}}=\hbar^{2}(k^{n,m}_{1}+k^{n,m}_{2})^{2}/(8m^{*})$, which is determined by the
 condition:
 \begin{eqnarray}
 J_{m-1}(k^{n,m}_{1}R)J_{m}(k_{2}^{n,m}R)+J_{m}(k^{n,m}_{1}R)J_{m-1}(k^{n,m}_{2}R)=0
 \label{eq:corralcond}
 \end{eqnarray}
 The solutions to Eq. (\ref{eq:corralcond}) which can have nonzero amplitude at the center of the circular billiard, the degenerate states $\Psi_{n,0}(\vec{r})$ and $\Psi_{n,1}(\vec{r})$,
 essential come in two types of eigenstates.  The first type
 occurs at energies $E_{\text{eff}}^{n}$ which are only about one to two meV smaller in energy than for the eigenstates $J_{0}(k_{n,0}|\vec{r}|)$ in the absence of spin-orbit coupling.  These states, although possessing some $J_{1}(k|\vec{r}|)\exp(\pm\text{i}\theta)$
 character, are mostly $J_{0}(k|\vec{r}|)$ in character, which leads to large peaks in the $\Delta LDOS(E_{\text{eff}},0)$ at slightly
 lower energies than the corresponding peaks in the absence of spin-orbit coupling.
The second type of eigenstate determined by Eq.
(\ref{eq:corralcond}) occurs at energies in between the
aforementioned energies.  These eigenstates, which are closely
related to the $m=\pm 1$ eigenstates in the absence of spin-orbit
coupling, $J_{\pm
1}(\overline{k}_{n}|\vec{r}|)\exp(\pm\text{i}\theta)$, possess a
small amount of $J_{0}(k|\vec{r}|)$ character due to spin-orbit
coupling, which can lead to new, albeit small, peaks in the $\Delta
LDOS(E_{\text{eff}},0)$ at these energies.  The small peak in the
$\Delta LDOS(E_{\text{eff}}=31 \text{meV},0)$ (and more clearly
shown in the inset in Fig. \ref{fig:fig2}(A)) corresponds roughly to
such an eigenstate, which, for the circular billiard with spin-orbit
coupling, has an energy of $E_{\text{eff}}=29.1$ meV.

Besides the small shift in the peaks of the $\Delta
LDOS(E_{\text{eff}},0)$ and the small peak at $E_{\text{eff}}=31$
meV, the observed difference in the $\Delta LDOS(E_{\text{eff}},0)$
with and without spin-orbit coupling is relatively small. However,
the $\Delta LDOS$ at other places inside the quantum corral can show
considerable differences when spin-orbit coupling is included.  A
slice of the $\Delta LDOS(E_{\text{eff}},\vec{r})$ through the
quantum corral is shown in Figs. \ref{fig:fig2}(B) and
\ref{fig:fig2}(C) for the (C) second peak in $\Delta
LDOS(E_{\text{eff}},0)$ [$E_{\text{eff}}=58.7$ meV (solid curve) and
$E_{\text{eff}}=60.4$ meV (dashed curve)]and for the (B) third peak
in $\Delta LDOS(E_{\text{eff}},0)$ [$E_{\text{eff}}=144.1$ meV
(solid curve) and $E_{\text{eff}}=145.8$ meV (dashed curve)], where
the black rectangles centered at $\pm 88.7 \AA$ correspond to the
positions of the adatoms in the slice. Note that in the simulations,
the $\Delta LDOS(E_{\text{eff}},\vec{r})$ is never calculated within
$6\AA$ of the adatoms.  As the electron bounces around in the
corral, it undergoes an effective spin rotation due to spin-orbit
coupling, which modulates the interference patterns seen in the
quantum corral, resulting in an enhancement (Fig. \ref{fig:fig2}(C))
or a decrease (Fig. \ref{fig:fig2}(B)) in the $\Delta
LDOS(E_{\text{eff}},\vec{r})$ near the edges of the corral.

\begin{figure}
\includegraphics*[height=14.7cm,width = 12.7cm]{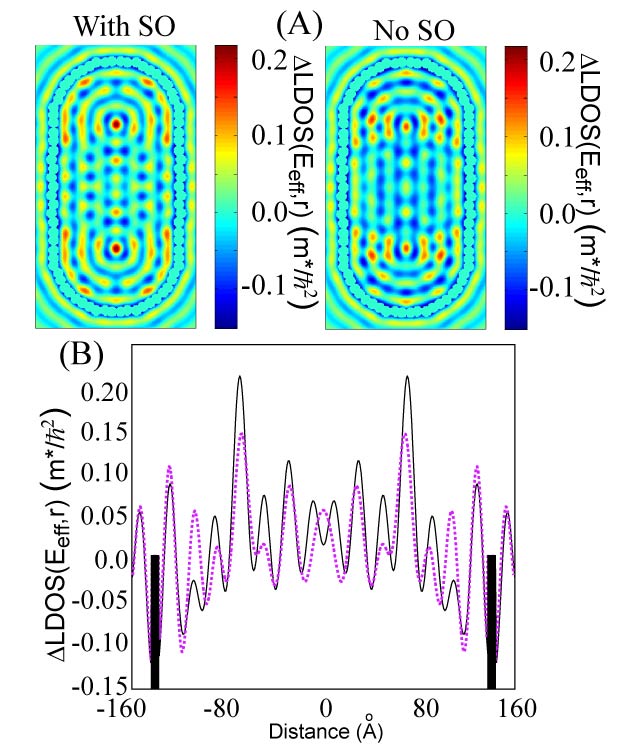}
\caption{(Color online) Simulation of the $\Delta
LDOS(E_{\text{eff}},\vec{r})$ for a 78 adatom quantum corral stadium
billiard of width $141\AA$ and length $285\AA$ at
$E_{\text{eff}}=410$ meV, with (left) and without (right) spin-orbit
coupling. Although the general features are similar, inclusion of
spin-orbit coupling can enhance or diminish features in the $\Delta
LDOS$ along with introducing additional peaks in the $\Delta LDOS$.
This can be more clearly seen in (B), where a slice through the
center of the stadium along its long dimension has been plotted with
spin-orbit (solid curve) and without (purple dashed curve)
spin-orbit coupling.  The black rectangles indicate the locations of
the adatoms through the slice. Besides differences in peak
intensity, a splitting of the $\Delta LDOS$ occurs at the center of
the stadium with spin-orbit coupling (peak to peak distance of
$\approx 18\AA$, which is absent when spin-orbit coupling is not
included.
 }\label{fig:fig3}
\end{figure}

Besides the circular corral, another corral simulated in this work
was a 78 adatom stadium billiard of dimensions $141\AA$ by $285\AA$,
where the adatoms were again placed atop a hexagonal lattice. Figure
\ref{fig:fig3} gives the $\Delta LDOS(E_{\text{eff}},\vec{r})$ with
and without spin-orbit coupling for roughly zero bias voltage
between the tip and surface, i.e., $E_{\text{eff}}=410$ meV.
Calculations performed at different $E_{\text{eff}}$ gave similar
results (data not shown).  As for the circular corrals, the $\Delta
LDOS(E_{\text{eff}},\vec{r})$ was artificially set to zero within
$6\AA$ of each adatom, which makes the adatom positions clearly
visible in Fig. \ref{fig:fig3}(A). Although the general structure of
the $\Delta LDOS(E_{\text{eff}},\vec{r})$ with and without
spin-orbit coupling appears similar, spin-orbit coupling causes
additional structure, such as splittings and intensity variations,
to appear in the $\Delta LDOS(E_{\text{eff}},\vec{r})$ as is shown
in Fig. \ref{fig:fig3}(A).  Such spin-orbit induced interference
effects can be more clearly seen in Fig. \ref{fig:fig3}(B), which
plots a slice of the $\Delta LDOS(E_{\text{eff}},\vec{r})$ through
the center of the stadium corral along the long dimension of the
corral. As with the circular corral, changes in the amplitude of the
$\Delta LDOS(E_{\text{eff}},\vec{r})$ are seen near the adatoms
(black rectangles in Fig. \ref{fig:fig3}(B)).  However, spin-orbit
coupling causes a splitting in the $\Delta
LDOS(E_{\text{eff}},\vec{r})$ near the center of the stadium, where
the peak to peak distance is roughly $18\AA$. Additional splittings
and modulations of $\Delta LDOS(E_{\text{eff}},\vec{r})$ can also be
seen in Fig. \ref{fig:fig3}(A).  These calculations clearly
demonstrate that spin-orbit coupling can generate significant
changes to the $\Delta LDOS(E_{\text{eff}},\vec{r})$ in quantum
corrals.

 In this work, we have examined the effects of spin-orbit coupling
 on the local density of states for quantum corrals formed atop the Au(111) surface.  Changes in the $LDOS(E_{\text{eff}},\vec{r})$ in both
 circular and stadium corrals indicate that spin-orbit induced interference effects should be visible using
 STM on the Au(111) surface, contrary to previous theoretical arguments\cite{Petersen00}.   The modulations in the $LDOS(E_{\text{eff}},\vec{r})$
 were a result of non-collinear multiple scattering trajectories, such as
those found in quantum corrals, which can generate an effective
spin-rotation in the presence of spin-orbit coupling.  Since the
previous experimental data on quantum corrals is quite good and can
be accurately described by multiple scattering
theory\cite{Heller94,Fiete03}, the predicted spin-orbit induced
interference in such systems should also be experimentally
observable.  Furthermore, this work can also be extended to the case
of quantum corrals comprised of magnetic adatoms, where, through the
spin-orbit coupling of the surface state electrons\cite{Heide07},
effective interactions between the magnetic adatoms can be
generated. The methodology in this paper can also be used to
calculate the $LDOS$ for superlattices formed from localized
structures.  In the future, ab initio calculations of STM images in
quantum corrals\cite{Stepanyuk05} with spin-orbit coupling will be
performed, in addition to examining the effect of the herringbone
reconstruction on the resulting $LDOS$ in quantum corrals on
Au(111).

JDW would like to thank Prof. Yung-Ya Lin for his support. This work
was supported by NSF NSEC.

\end{document}